# Detection chamber for search of low Mass WIMP and solar axions

### B. M. Ovchinnikov*, V. V. Parusov

Institute for Nuclear Research, Russian Academy of Sciences, Moscow, Russia
*Corresponding Author: ovchin@inr.ru

### Abstract

A chamber for direct detection of WIMPs with masses < 10 GeV/$c^2$ and solar axions is developed. The chamber is filled with a gas mixture of $H_2$ +3ppm TMAE (1,5,10 bar), or $D_2$ + 3ppm TMAE. These gas fillings make it possible to suppress an electron background. For detection of events, the system GEM+pin-anode with a multiplication coefficient of about $10^5$ is constructed.

## 1. Introduction

Currently an existence of the Dark Matter of the Universe is confirmed by multiple astronomical observations. Because the WIMP masses > 10 GeV in laboratory experiments are not discovered, it is necessary to search the WIMP with masses ≤ 10 GeV. J.Va'vra [1] have supposed, that yearly modulation effect in experiments [2] is explained by low mass WIMP scattering on protons of $H_2O$ molecules which contamination in detector NaI [2]. In work [5] it is shown, that the lightest neutralino is the best candidate for the Dark Matter WIMP.

## 2. Methodology of Search for low Mass WIMP.

In this work for the search of low mass WIMPs it is proposed to use a detection chamber [3] with $H_2$ + 3ppm TMAE gas mixture filling.

Collisions of WIMPs with $H_2$ provide recoil protons with energies of several keV (see Table 1). The addition of TMAE with a low ionization potential (5,36eV) provides detection of the recoil protons. As another filling of the chamber it can be used a mixture of deuterium + 3ppm TMAE, which provides a number of advantages. First, the energy of the D-recoil is twice larger than the proton recoil. Second, the nuclear spin of deuterium is equal to 1, while hydrogen has a nuclear spin of ½, which provides three times larger collisional cross section, compared to hydrogen [6].

The $H_2$-filling provides an efficient suppression of the electron background, because of the short track of recoil protons, compared to one of background electrons [4].

For investigation of the $H_2$ +3ppm TMAE gas mixtures, the chamber with GEM +pin-anode detection system was constructed (Fig 1.). The chamber was filled with $H_2$ + 3ppm TMAE under pressures of 1, 5 or 10 bar. The results are shown in Fig 2. The measurements show that the TMAE-addition leads to stabilization of the propotional discharge in the detection system.

The chamber is placed in a low background laboratory with additional shielding for search of the yearly or daily modulation effects.

## 3. Search of axions emitted by Sun.

For search of solar axions we used $D_2$ + 3ppm TMAE filling of the chamber.

As well as the mass and energy of axion are equal to ~ 1 keV [7], it transfers the energy to recoil deuterium. This makes it possible to suppress the background.

Table 1. Calculated maximum nuclear recoil energy EkeVnr as a function of WIMP mass for two targets: hydrogen and sodium (H ,Na) [1].

| WIMP mass [GeV/c²] | Nucleus | $E_{keVnr}$[keV] |
|---|---|---|
| 0.5 | *H* | 1.91 |
| 1.0 | *H* | 4.30 |
| 1.5 | *H* | 6.20 |
| 2.0 | *H* | 7.65 |
| 2.5 | *H* | 8.78 |
| 3.0 | *H* | 9.68 |
| 0.5 | *Na* | 0.19 |
| 1.0 | *Na* | 0.73 |
| 1.5 | *Na* | 1.57 |
| 4.0 | *Na* | 9.07 |

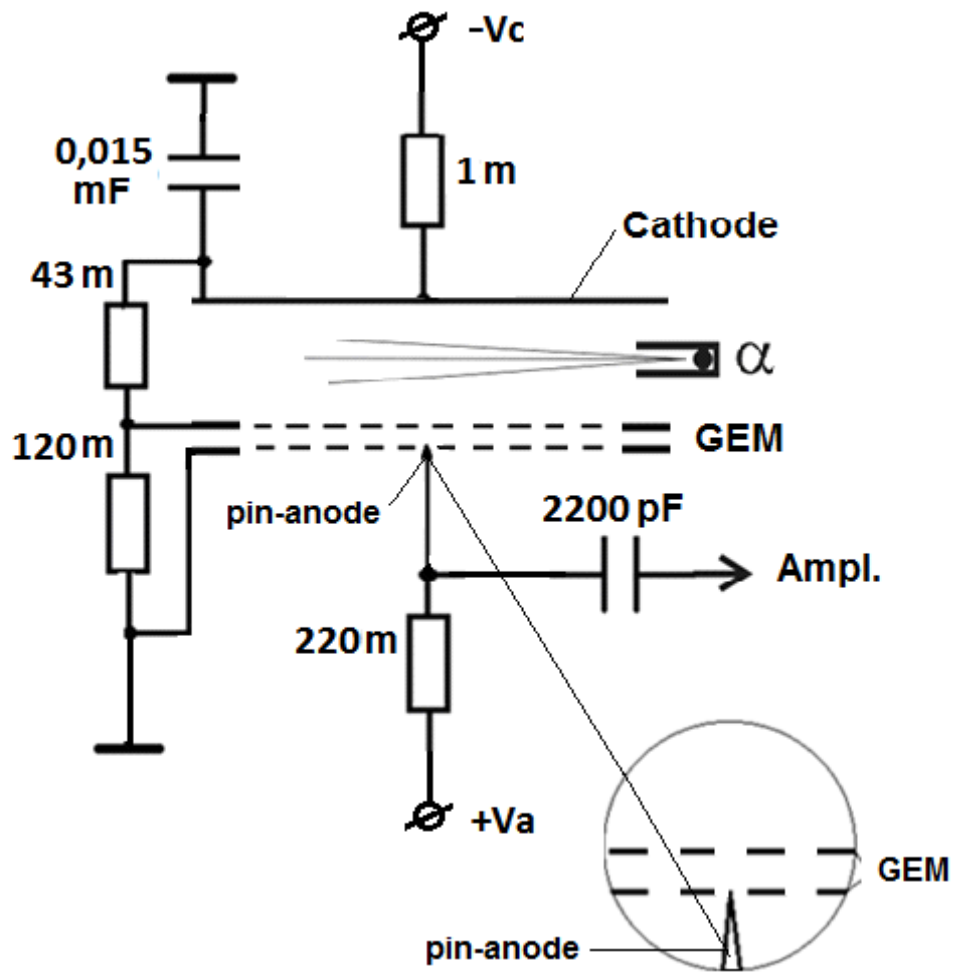

**Figure 1.** The "metallic GEM + pin-anode" detection system for investigation of $H_2$ + 3ppm TMAE gas filling mixture.

+Va=0÷3200V, -Vc=580V(5bar), -Vc=860V(10bar).

## 4.Results

The dependence of the electron multiplication factor ($K_{ampl}$) from the potential on pin-anode ($V_{anode}$) is shown in Fig. 2.

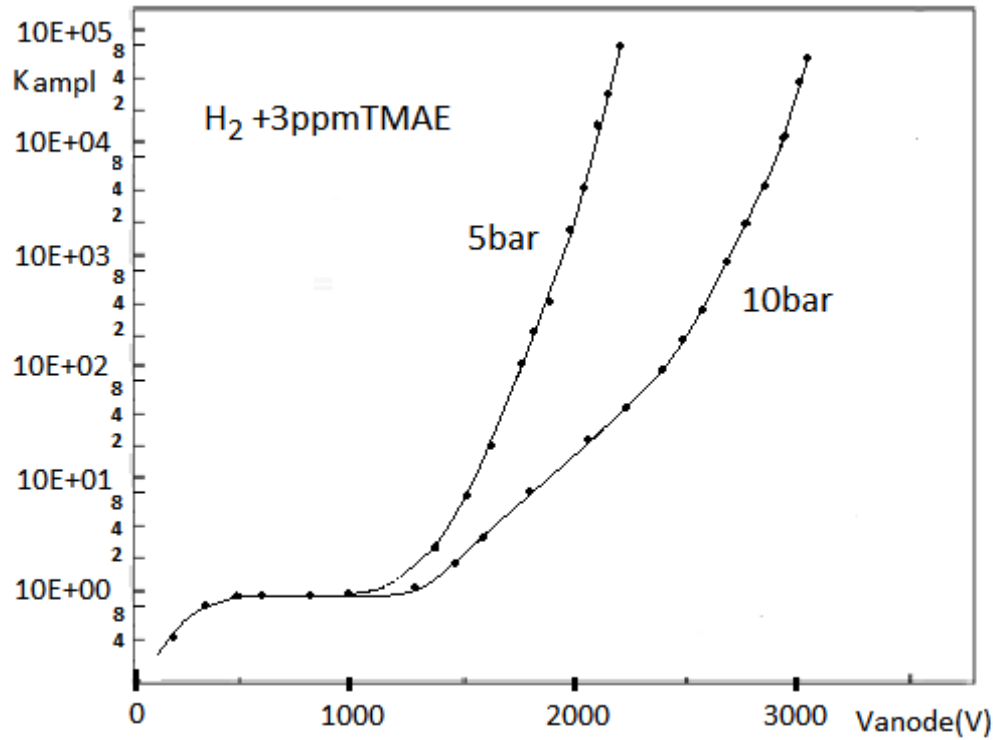

**Figure 2.** The dependence of $K_{ampl.}$ from the potential on pin-anode.

The TMAE-addition stabilizes the propotional discharge in detection system.

($K_{ampl}= Q_{ampl}/Q_{ioniz}$, where $Q_{ampl}$ –the charge detected, $Q_{ioniz}$ –the ionization charge).

## 5.Discussion

The hydrogen and deuterium fillings of the detection chamber make it possible to detect WIMP with low mass and axion, as well as to suppress the electron background.

## 6.Conclusion

The next steps will include the development of an electronic system for detecting of events in the detection chamber and construction of the low background shielding in the low background laboratory.

## 7. Acknowledgement

The autors would like to thank Dr. I.I.Tkachev for many useful discussions.